\documentclass[5p,times]{elsarticle}
\usepackage{float}
\usepackage{amssymb}
\usepackage{fixltx2e}

\usepackage{hyperref}
\hypersetup{
	colorlinks=true,
	urlcolor= blue,
	citecolor=blue,
	linkcolor= blue,
	bookmarks=true,
	bookmarksopen=false,
}

\biboptions{numbers,sort&compress}

\journal{}
\date{December, 2019}

\begin{document}

\begin{frontmatter}

\title{Antimonene/Bismuthene Vertical Van-der Waals
	Heterostructure: A Computational Study}

\author[1]{Shobair Mohammadi Mozvashi}

\author[2]{Sahar Izadi Vishkayi}

\author[1]{Meysam Bagheri Tagani\corref{cor1}}
\ead{m{\_}bagheri@guilan.ac.ir}

\address[1]{Department of physics,  Computational Nanophysics Laboratory (CNL),
University of Guilan, P.O.Box 41335-1914, Rasht, Iran}

\address[2]{School of Nano Science, Institute for Research in Fundamental
	Sciences
	(IPM), P. O. Box 19395-5531, Tehran, Iran}

\cortext[cor1]{Corresponding author}

\begin{abstract}
In this paper, the  structural, electronic, mechanical and optical
properties of
Antimonene/Bismuthene Van-der Waals heterostructure (Sb/Bi HS) were
calculated based on the first principle density functional theory. We explored
different stacks of Sb/Bi HS to find the most, and the least stable staking for
this heterostructure. At the GGA level
of theory, the most stable model is a semiconductor with an indirect
bandgap of 159 meV. However, when the spin-orbit (SO) interaction is
considered, the VBM
and CBM touch the Fermi level and the HS becomes a semimetal. Our results also
show
that the electronic properties of the HS  are
robust against the external electric field and biaxial strain. Young's modulus
was calculated of 64.3 N/m which predicts this HS as a resistant material
against being stretched or compressed. The calculated optical properties,
similar to monolayer Antimonene, are completely dependent on the
polarization
of incident light and differ when parallel or perpendicular polarization is
considered. Moreover, the absorption coefficient of Sb/Bi HS for perpendicular
polarization in the visible region is significantly increased in comparison
with the
monolayer Antimonene. High structural stability, electronic and
mechanical robustness against electric field and strain, along with
polarization-dependent optical properties of this HS, promise for
its applications in beam splitters and nano-scale mirrors.
\end{abstract}

\begin{keyword}
DFT \sep 2D materials \sep Antimonene \sep Bismuthene \sep
Heterostructure
\end{keyword}

\end{frontmatter}

\section{Introduction}

Two dimensional (2D) materials, due to their unique
properties, such as high
carrier mobility \cite{lu14}, superior mechanical characteristics
\cite{frank07}, and high ratio between their lateral size (1-1000 $\mu$m) and
thickness ($<$ 1nm) \cite{novoselov2012roadmap}  have attracted a great deal of
attention in the last decade, and annually this attraction becomes even
greater.
The discovery of Graphene and its unique properties in 2004 \cite{novoselov04},
opened a road to new investigations and observations in 2D materials
because, before that, it assumed that 2D materials are thermodynamically
unstable \cite{geim11}. After that, several types of 2D materials theoretically
investigated, and/or experimentally exfoliated, including hexagonal boron
nitride, Transition metal dichalcogenides, perovskites, and other elemental 2D
materials such as group IV and V monolayers \cite{kamal15,
	zhang2015,ni11,davila14,chhowalla13,liu03, bafekry2019, mohebpour2018}.
	Because
of various
applications,
such as sensors \cite{xu2018two}, field effect transistors
\cite{reddy2011graphene}, light-emitting diodes \cite{erchak01}, solar cells
\cite{saffari2017dft}, and energy-conserving devices \cite{cai10} in nano-scale
size, the research interest in 2D materials remains attractive for many years.

Antimonene and Bismuthene, from group V elemental monolayers, are relatively
novel members of 2D materials which their electronic,
mechanical and optical properties have been investigated in numerous
experimental and theoretical studies in recent years  \cite{wang15,
	akturk15,akturk16,gibaja16,singh16,xu17,guo2018biaxial}. The studies imply
	that
these two monolayers are both stable in $ \beta $ (buckled hexagonal) phase and
predict an
indirect bandgap for Antimonene and a direct narrow bandgap for Bismuthene.
Besides, it is predicted that the electronic properties of Antimonene are
sensitive to strain and it is possible to tune its bandgap in this way. Wang et
al. predicted that by applying a tensile biaxial strain of ~5$\%$ the bandgap
of Antimonene becomes direct and increases to its maximum value \cite{wang15}.

On the other hand, nowadays designing and investigating the Van-der Waals
heterostructures has turned into an interesting method for broadening the scope
of
research in the field of 2D materials. Scientists can simply predict and/or
observe very interesting properties by combining
already discovered materials, not discovering new ones. For instance, Lu et
al. by designing heterostructures of Antimonene and three 2D materials
(Graphene, Arsenene, and h-BN), could predict heterostructures with tunable
bandgaps
between 0 to 1 eV which have potential applications in near and mid-infrared
detectors \cite{lu16}. Meng et al. reported that one can enhance the efficiency
of MoS$_{2}$ based solar cells, by 5.23$\%$, using a MoS$_{2}$/Si
heterostructure \cite{tsai14}. Also Chen et al. predicted the opening of a
direct and
tunable bandgap of ~400 meV in Germanene, by designing an Antimonene/Germanene
Van-der Waals heterostructure \cite{chen16}.

Our purpose in this study is to apply a tensile strain of 5$\%$ to the
Antimonene, by designing an Antimonene/Bismuthene Van-der Waals
heterostructure. One of the natural methods for applying a biaxial strain is to
form a material on a substrate with a different lattice constant. If the
substrate lattice constant is larger or smaller than the superstrate, a tensile
or compressive strain applies to the latter, respectively. Reviewing the
literature, it raises that
the lattice constant for Bismuthene is $\sim$5$\%$ larger than
Antimonene's.
Therefore,
by the case of forming an Sb/Bi HS, with fixed Bismuthene substrate, a tensile
strain
of about 5$\%$ would be applied to the Antimonene superstrate. In this paper,
we
intend to probe if it is theoretically possible to design a stable Van-der
Waals heterostructure with such mismatching monolayers. After investigating the
structural
characteristics and testing the stability, we look into the electronic,
mechanical and optical properties of Sb/Bi HS and discuss its potential
applications.

\section{Computational Details}
For all the simulations, the Spanish package solution, SIESTA
\cite{soler2002siesta} was employed which is based on self-consistent density
functional theory (DFT) and standard pseudopotentials. The exchange-correlation
interactions were estimated through generalized gradient approximation (GGA),
with parametrization of Perdew, Burke, and Ernzerhof (PBE) \cite{perdew97}. For
the electronic and optical properties of the heterostructures, the spin-orbit
coupling (SOC) was also considered in addition to GGA (SOGGA). In all the
calculations, the reciprocal space was sampled by a mesh of
31$\times$31$\times$1 k points in the Brillouin zone and the density mesh
cut-off was set on 100 Ha. For considering the Van-der Waals interactions
between layers, the DFT-D2 correction of Grimme \cite{grimme04}
was used. A vacuum space of 20 \r{A} was considered in the z-direction to
prevent unwanted interactions.  Besides, all the pristine structures were
relaxed to a force and stress of 0.001 eV/\r{A} and 0.001 GPa, respectively, but
in the strain related investigations the minimization of stress was canceled.

\section{Results and Discussion}
\subsection{Monolayer Antimonene and Bismuthene}
\begin{figure*}[h]
	\centering
	\includegraphics[width=0.9\textwidth]{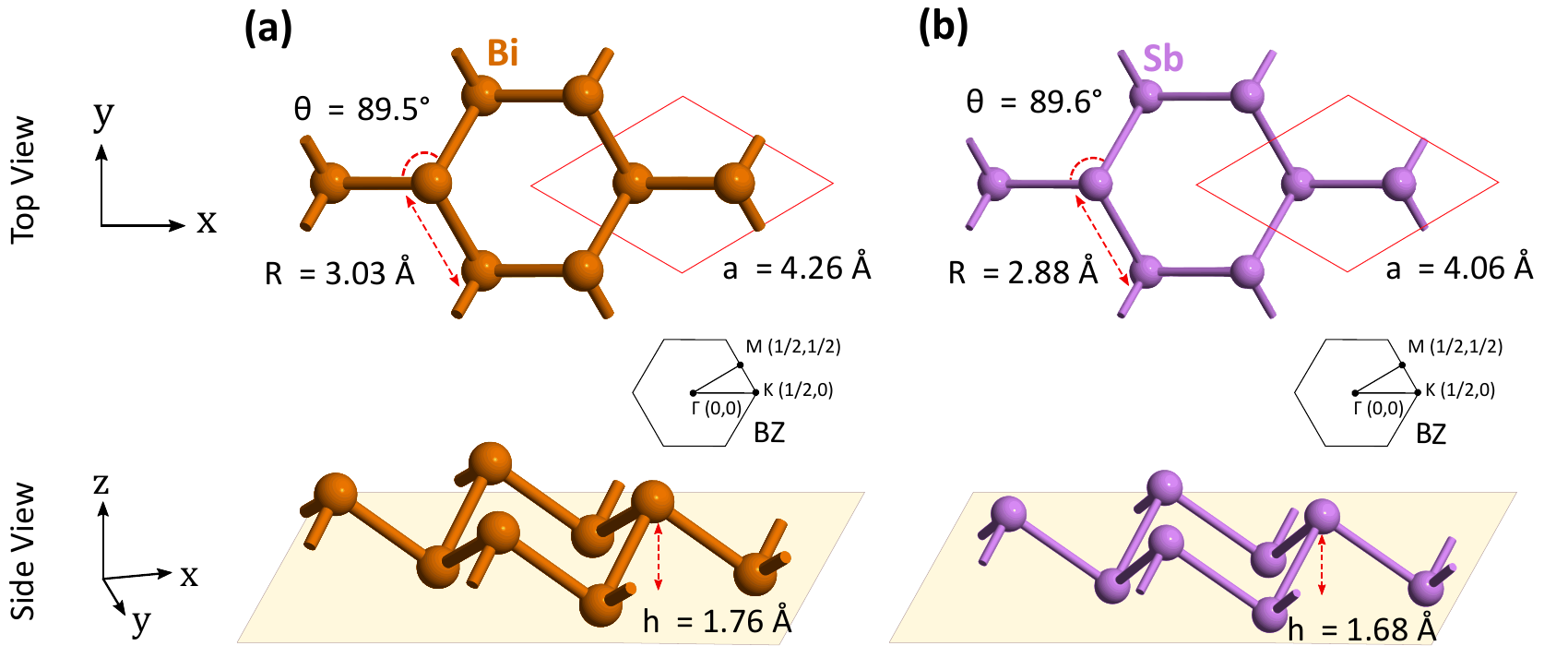}
	\caption{Structural configurations of monolayer
		Bismuthene
		(\textbf{a}), and Antimonene (\textbf{b}). unit cells, atomic bonds
		and Brillouin zones are depicted in the figure.}
	\label{config}
\end{figure*}

Our quest in this part is not for expressing a novel and independent
assessment about these monolayers, but only for validating our methods, and
verifying prior studies. However, before bringing up the discussion about Sb/Bi
HS, it is needed to investigate and review the properties of each monolayer
separately.
The pristine structures of Antimonene and Bismuthene are buckled hexagonal,
where the atoms are seen like a hexagon from the top view, but they are
not in the same plane plane, that is, the bonding atoms have a distance with
each other in
the
z-direction, called ``buckling" (Fig.\ref{config}). Previous studies have
calculated the lattice constant from 3.94 to 4.12 \r{A} for Antimonene
\cite{zhang2015,akturk15,wang15} and 4.30 to 4.39 \r{A} for Bismuthene
\cite{mounet18,zhang16,liu17}. We calculated the Antimonene and
Bismuthene lattice constants as 4.06
and 4.26 \r{A}, respectively. Regarding the variety of approximations,  basis
sets, pseudopotentials and code packages, this order of difference in results
is inevitable, thus we assess our calculations rational and reliable.

\begin{table}[h]
	\scriptsize
	\caption{Structural parameters of monolayer
		Bismuthene
		and Antimonene. lattice constants (a), atomic bonds (R), buckilng
		heights
		(h), angles between two bonds ($\theta$), cohesive energies ($E_c$),
		and bandgaps ($E_g$) are shown and compared with previous studies.}
	\label{configtab}
	
	\begin{tabular}{lllllcc}
		\hline\hline
		& a (\AA) & {R (\AA)} & {h (\AA)} & {$\theta$ ($^\circ$)} & $E_c$
		(eV/atom) &
		$E_g$, GGA (eV)\\
		
		\hline
		\textbf{Sb}\\
		This study &{4.06}&{2.88}&{1.68}&{89.6}&{-3.14}&{1.24 (ind)}\\
		\cite{akturk15}&{4.04}&{2.87}&{1.67}&{89.0}&{-2.87$^{\rm a}$}&{1.04
		(ind)}\\
		\cite{wang15}&{4.12}&{2.89}&{1.64}&{90.8}&{-4.03}&{0.76 (ind)}\\
		\cite{liu19}&{4.08}&{2.88}&{1.66}&{90}&{n/a}&{n/a}\\
		\hline
		\textbf{Bi}\\
		This study &{4.26}&{3.03}&{1.76}&{89.5}&{-3.00}&{0.75 (dir)}\\
		\cite{mounet18}&{4.30}&{3.03}&{1.75}&{90.2}&{n/a}&{0.60 (dir)}\\
		\cite{zhang16}&{4.34}&{n/a}&{1.73}&{n/a}&{n/a}&{0.99 (dir)}\\
		\cite{liu17}&{4.39}&{3.07}&{1.73}&{91.2}&{-3.04$^{\rm a}$}&{0.46
		(dir)$^{\rm
				b}$}\\
		\hline\hline
	\end{tabular}
	\\
	\\
	\scriptsize $^{\rm a}$ A conversion of unit or sign was required.\\
	\scriptsize $^{\rm b}$ GGA + SOC
\end{table}

The structures were relaxed and their cohesive energies were calculated with the
equation below,
\begin{equation} \label{cohesive}
E_c=\frac{E_{sheet}-2E_{atom}}{2}
\end{equation}

\noindent where, $ E_{sheet} $ is the total energy of each monolayer, and
$E_{atom}$ is the energy of each isolated Sb or Bi atoms considering
spin polarization. Based on the equation (\ref{cohesive}), the value for
cohesive energy should be negative for stable structures. Generally, our
results about lattice constants, bond lengths, bond angles, buckling heights,
cohesive energies and bandgaps are shown and compared
with previous studies in Table \ref{configtab}, which are in a good
agreement with the literature.

\begin{figure*}[h]
	\centering
	\includegraphics[width=0.9\textwidth]{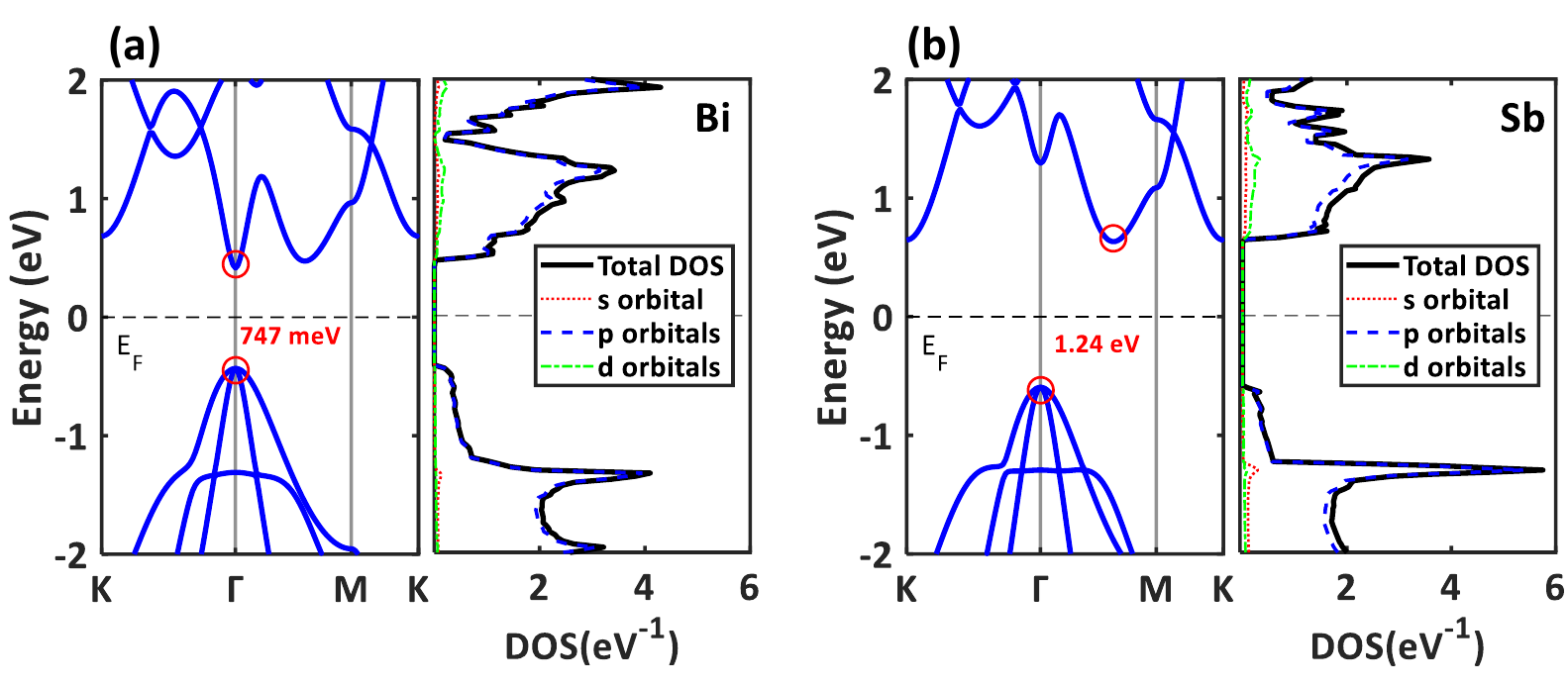}
	\caption{Band structures and density of states of
		Bismuthene (\textbf{a}), and Antimonene (\textbf{b}). Fermi level is
		shifted to zero.}
	\label{fig2}
\end{figure*}

Fig.\ref{fig2} represents the band structures and partial density of states
related to
both monolayers in the pristine configurations. The band structures show a
direct bandgap of 847 meV for Bismuthene and
an indirect bandgap of 1.24 eV for Antimonene. The density of states represent
that in both monolayers, the p orbitals are most
responsible for electronic properties, and the s and d orbitals do not have any
significant contribution.

\begin{figure*}[h]
	\centering
	\includegraphics[width=0.9\textwidth]{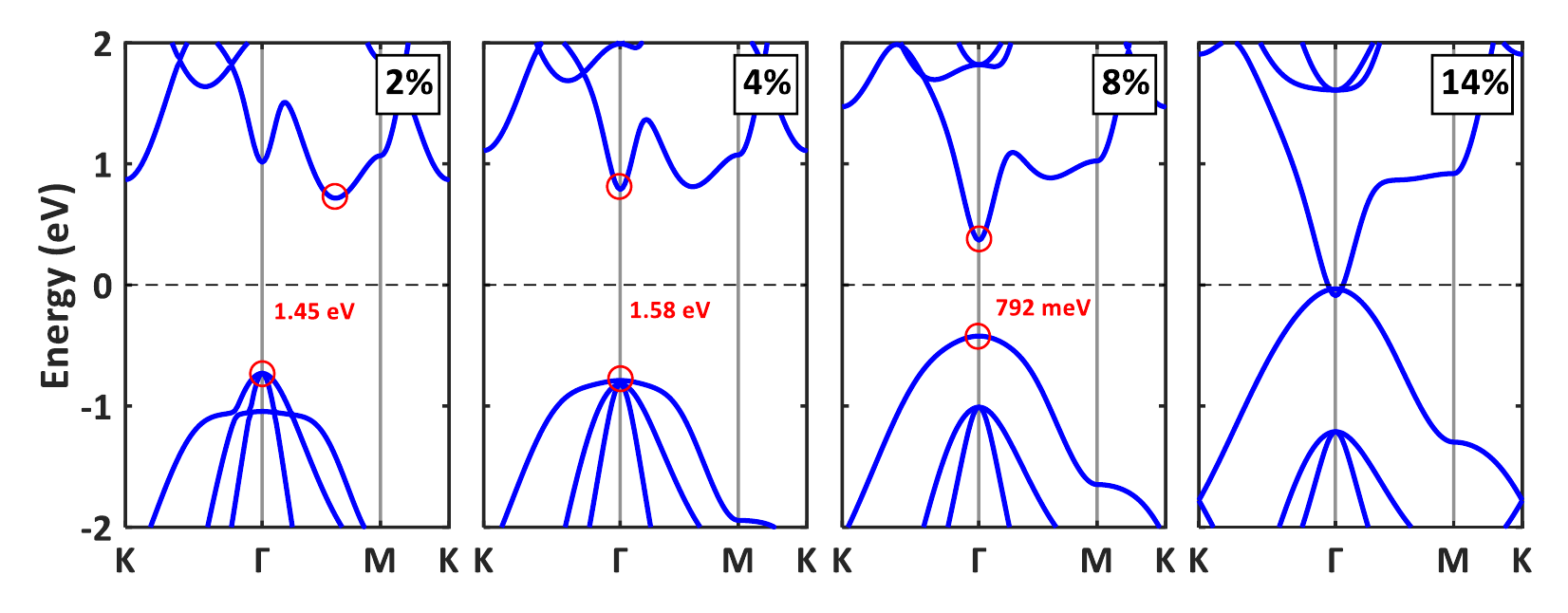}
	\caption{Variation of the GGA band structure of  Antimonene
		with tensile strain. Each sub-figure is related to the tensile strain
		depicted at up-right text boxes.}
	\label{fig3}
\end{figure*}

\begin{figure}[h]
	\centering
	\includegraphics[width=0.4\textwidth]{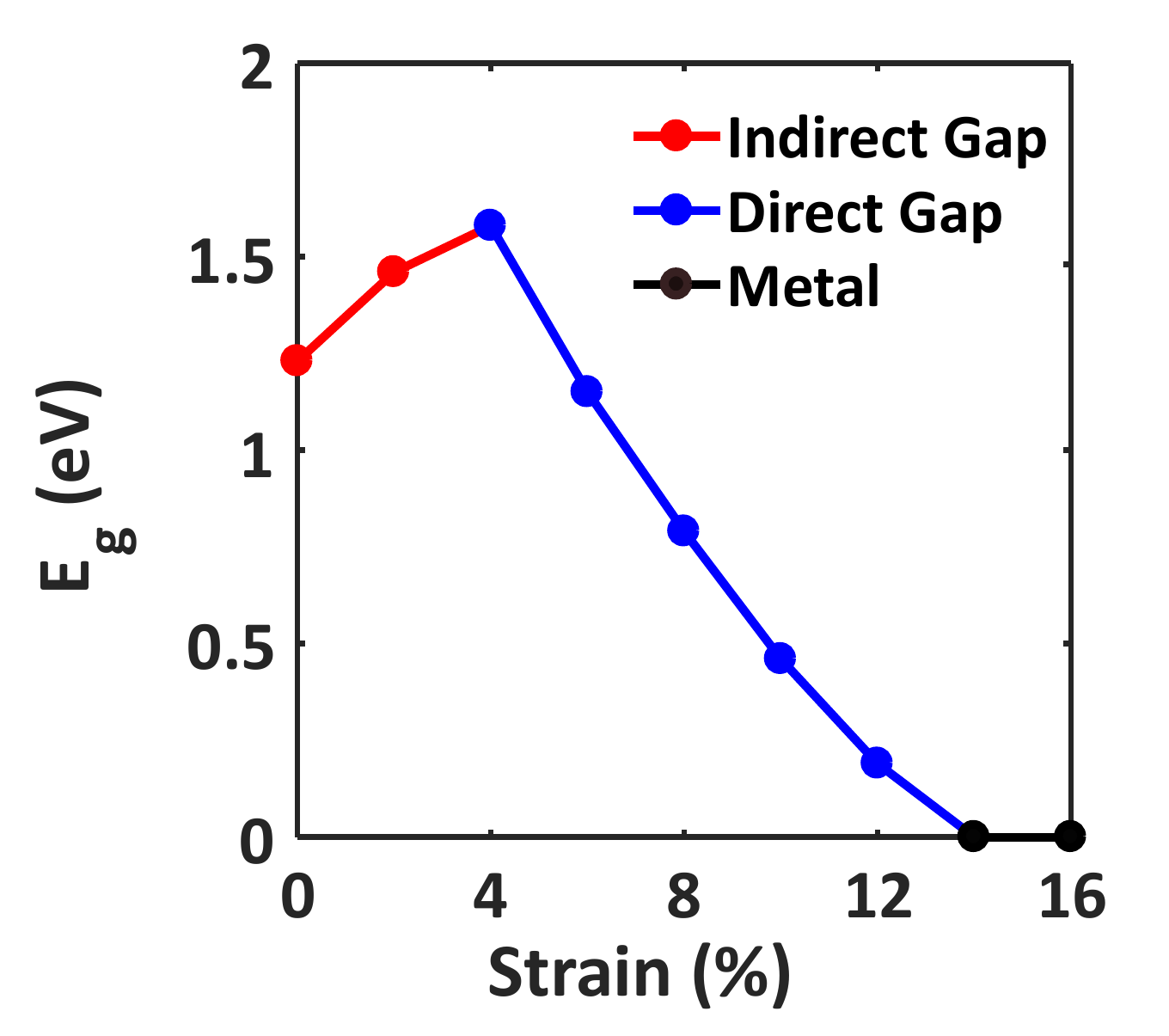}
	\caption{Variation of monolayer Antimonene GGA bandgap
		with tensile strain.}
	\label{fig4}
\end{figure}

As mentioned above, previous studies have noted that
electronic properties of Antimonene are sensitive to strain and an
indirect to direct transition would occur under the tensile biaxial strain of
5$\%$
\cite{wang15, zhao15}. Because our purpose in this study is to probe the
behavior of Antimonene under substrate generated strain, we also investigated
the effect of strain to have a comprehensive review, and re-test
our computational method. The biaxial strain is defined by the equation below,
\begin{equation} \label{strain}
\epsilon = \frac{(a-a_0)}{a_0}
\end{equation}
\noindent where  $a_0$ and $a$ are lattice constants in pristine and
strained structures, respectively. Our results show that with short biaxial
tensile strain ($<$4$\%$) the bandgap of Antimonene increases. It reaches to
the maximum
value
of 1.58 eV under the strain of 4$\%$, and an indirect to direct transition
occurs at
this point. After that the bandgap decreases softly and finally under the
strain of 14$\%$ it is closed and the material turns into a semimetal (Figs.
\ref{fig3} and \ref{fig4}). The behavior we predict agrees with the
previous reports. Good agreement of our results about the
monolayers with the literature, especially the prediction of the indirect to
direct transition,
makes us to trust our  method and gain sufficient motivation to continue our
investigations.

\subsection{Structural Properties of Antimonene/Bismuthene Heterostructure}
\begin{figure*}[h]
	\centering
	\includegraphics[width=0.9\textwidth]{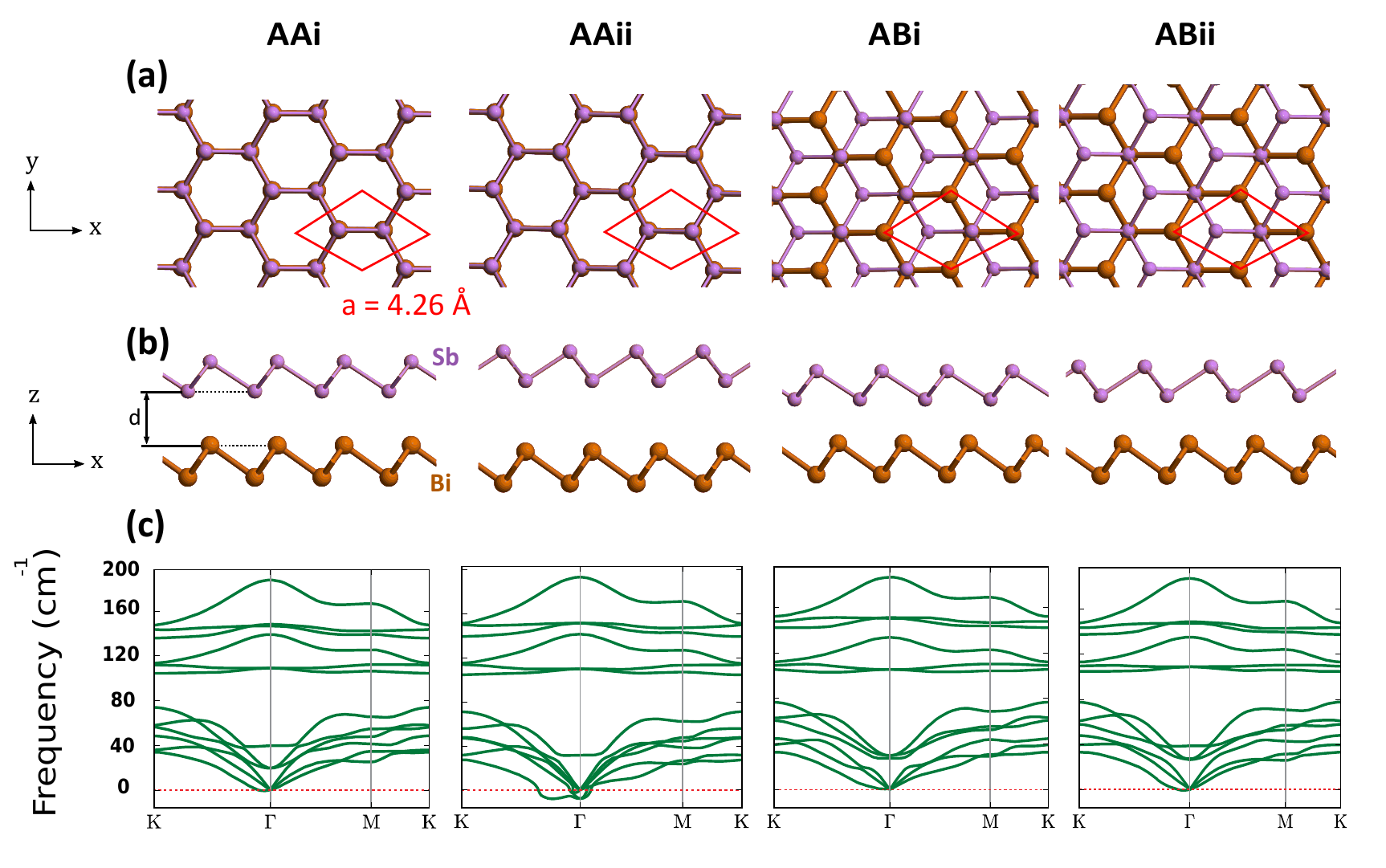}
	\caption{Four models of Sb/Bi HS.
		\textbf{(a)} The top view, \textbf{(b)} the side view, and \textbf{(c)}
		the phonon dispersion
		for each model.}
	\label{stackings}
\end{figure*}

As mentioned above, the lattice constants for Antimonene and
Bismuthene are 4.06 and 4.26 \AA, respectively. Therefore, these two
monolayers
have a mismatch of 4.93$\%$ with each other. In this study, we considered the
Bi layer as a fixed substrate and let the Sb layer to relax on it. Therefore a
tensile strain of 4.93$\%$ applies to the latter. We predicted four possible
models for the HS. AAi, AAii, ABi, and ABii. The AAi and ABi models are the
regular AA and AB stackings, while in AAii and ABii models the upper and lower
layers have different wrinkling directions. For example in the AAi model, the
upper Sb atom is on top of the upper Bi atom, while in the AAii model, the
upper Sb atom is on top of the lower Bi atom. These four types of stackings are
declared in Fig.\ref{stackings}a,b. In all
these
models, the lattice constant is 4.26 \AA\ which is equal to monolayer
Bismuthene's.

\begin{table*}[h]
	\centering
	\scriptsize
	\caption{Structural parameters of four possible models of
		the Sb/Bi HS: Sb-Sb and Bi-Bi bond lengths, inter-layer distances
		(d),
		cohesive energies ($E_c$), stress($\sigma$), and GGA bandgaps ($E_g$).}
	\label{stacktab}
	
	\begin{tabular}{lcccccc}
		\hline\hline
		Model & Sb-Sb ({\AA}) & Bi-Bi ({\AA}) & d ({\AA}) & $E_c$
		(meV/atom)
		& $\sigma$ (N/m) &$E_g$ (eV)\\
		\hline
		AAi & 2.94 & 3.03 & 2.98 & -155 & 1.81 & 356 (ind)\\
		AAii & 2.94 & 3.03 & 3.92 & -112 & 2.30 & 390 (dir)\\
		ABi & 2.93 & 3.03 & 2.48 & -175 & 0.28 & 159 (ind)\\
		ABii & 2.94 & 3.03 & 2.69 & -173 & 1.29 & 252 (ind)\\
		\hline\hline
	\end{tabular}
\end{table*}

\begin{figure}[h]
	\centering
	\includegraphics[width=0.4\textwidth]{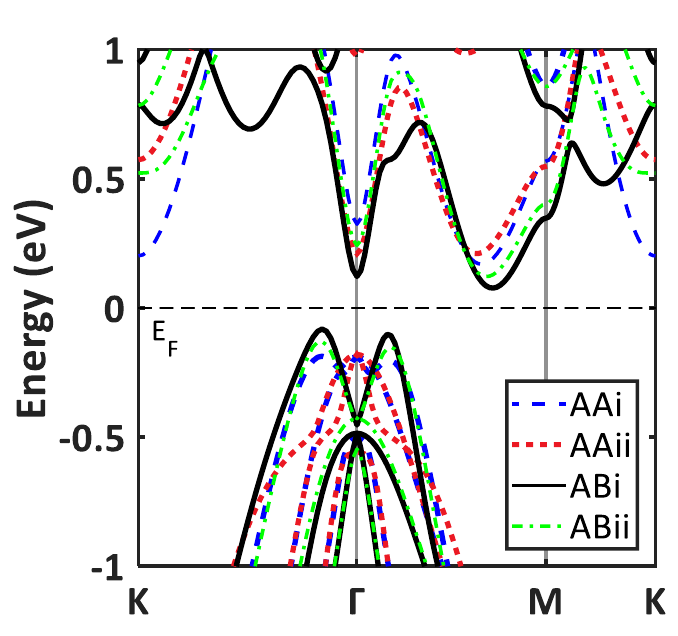}
	\caption{GGA band structures of the four models of
		Sb/Bi HS.}
	\label{bands}
\end{figure}

\noindent For all the models, the Sb-Sb and Bi-Bi bonds are $\sim$2.94
and 3.03 \AA, respectively, therefore the bond lengths are independent of
stacking
and there is a little recombination between different models.
The phonon dispersions (Fig.\ref{stackings}c)
show that three of the stackings are vibrationally stable, except Model
AAii in which the imaginary modes near
$\Gamma$ point show its instability. It should be noted that For
calculating the cohesive energies between layers, we used the equation below,
\begin{equation}
E_c=\frac{E_{HS}-(E_{Sb}+E_{Bi})}{4}
\end{equation}
\noindent where, $E_{HS}$ is the energy of the heterostructure, and $E_{Sb}$,
$E_{Bi}$ are energies of Sb and Bi layers, respectively. The largest and the
smallest absolute value of cohesive energy belongs to ABi (-175 meV/atom) and
AAii (-112 meV/atom) models, respectively, therefore, in terms of
cohesive Energy, the ABi and AAii stackings are the most and the least stable
models. According to Table \ref{stacktab}, the shortest inter-layer
distance (d) belongs to the ABi (2.48 {\AA}) and the longest one is for AAii
(3.92{\AA}) models, respectively. The shorter inter-layer distance means
that the interactions and forces are more intense between the layers and they
tend to attract each other more. Therefore the ABi model turns out more
favorable in this case too.  Also, surface
stress in ABi model is much
smaller in contrast with other models, while it is the highest in model AAii.
In summary, it could be said that the most stable model is ABi, while AAi and
ABii models are less stable and AAii model is completely unstable. Our
calculations in phonon
dispersion, cohesive energy, inter-layer distance, and stress among these four
models are supportive of each other.
By reviewing the previous works, it raises that the
model we call ABi was usually more stable in other Antimonene based
heterostructures as well. For example, Chen et al reported that in
Antimonene/Germanene heterostructure, a so-called ABI model is the most stable
stacking with a -283 meV/atom of cohesive
energy \cite{chen16}. Also, in Sb/InSe heterostructure, the most stable model
is reported a so-called H3 model with a cohesive energy of -23.86 eV/{\AA$^2$}
(by
multiplying by the unit cell area and dividing into the number of atoms, the
converted cohesive energy is gained -57.61 meV/atom.) which is similar to
our ABi model. Because of the high stability of ABi model, we continue our
investigations only with ABi model. From now on, when we refer to Sb/Bi HS, we
mean the ABi model of the Sb/Bi heterostructure.

\subsection{The Electronic properties of the Sb/Bi HS}
\subsubsection{The pristine structure}
\begin{figure*}[h]
	\centering
	\includegraphics[width=0.9\textwidth]{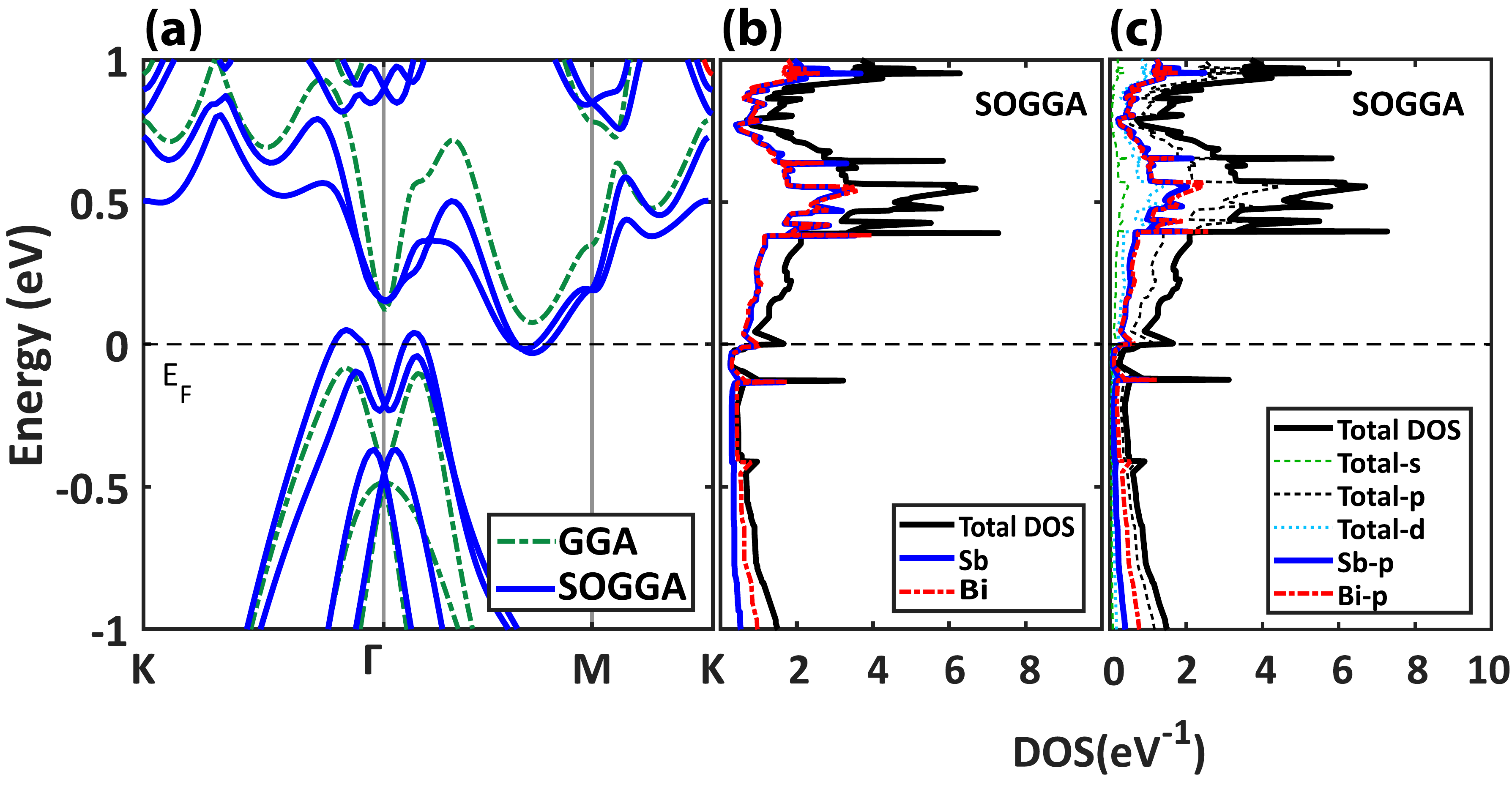}
	\caption{ The Electronic properties of the Sb/Bi HS.
		\textbf{(a)} Band
		structure at GGA and SOGGA levels of
		theory. \textbf{(b,c)} Atom and orbital projected density of states,
		respectively, both at
		SOGGA
		level of theory.}
	\label{pdos}
\end{figure*}

The calculated band structure for Sb/Bi HS (Fig.\ref{pdos}) shows that this
material is a semiconductor with an indirect bandgap of 159 meV at GGA and a
semimetal at GGA+SOC (SOGGA) level of theory. According to the heavy atoms
consisting the HS, especially Bismuth, the determinant influence
of SOC interactions are expectable. Generally, considering the SOC effect is
important in structures with heavy
elements \cite{manzeli20172d}, but little and neglectable for lighter ones
\cite{ezawa2017triplet}. As we know, considering the SOC eliminates
the energy degeneration and assets the electronic properties more precisely.
Therefore, it is expected that in case of potential experimental research, the
results would be more close to our SOGGA solutions. Therefore,  in all of our
electronic and optical
calculations, we continued at the SOGGA level of theory.

It can be seen from the atom projected density of states (Fig.\ref{pdos}b)
that in the valance band the contribution of Bi atoms is slightly higher than
Sb
ones but in the conduction band, they play the same role in the electronic
properties. Besides, the orbital projected density of
states (Fig.\ref{pdos}c) represents that the contribution of p orbitals
is
determinant and other orbitals have neglectable roles. Considering that both
consisting elements are in group V, the influence of p orbitals is natural.
Within  the whole showed range in the conduction band, and near the Fermi level
in the valance
band, the Sb and Bi p orbitals have a superposition, therefore the mechanism in
which the layers connect to each other may be through orbital hybridization.
There are
similar reports about the superposition of orbital projected density of states
in Van-der Waals heterostructures, and existence of orbital hybridization
between the layers; such as Antimonene/Arsenene and
Antimonene/Germanene \cite{lu16, chen16}.

\begin{figure*}[h]
	\centering
	\includegraphics[width=0.9\textwidth]{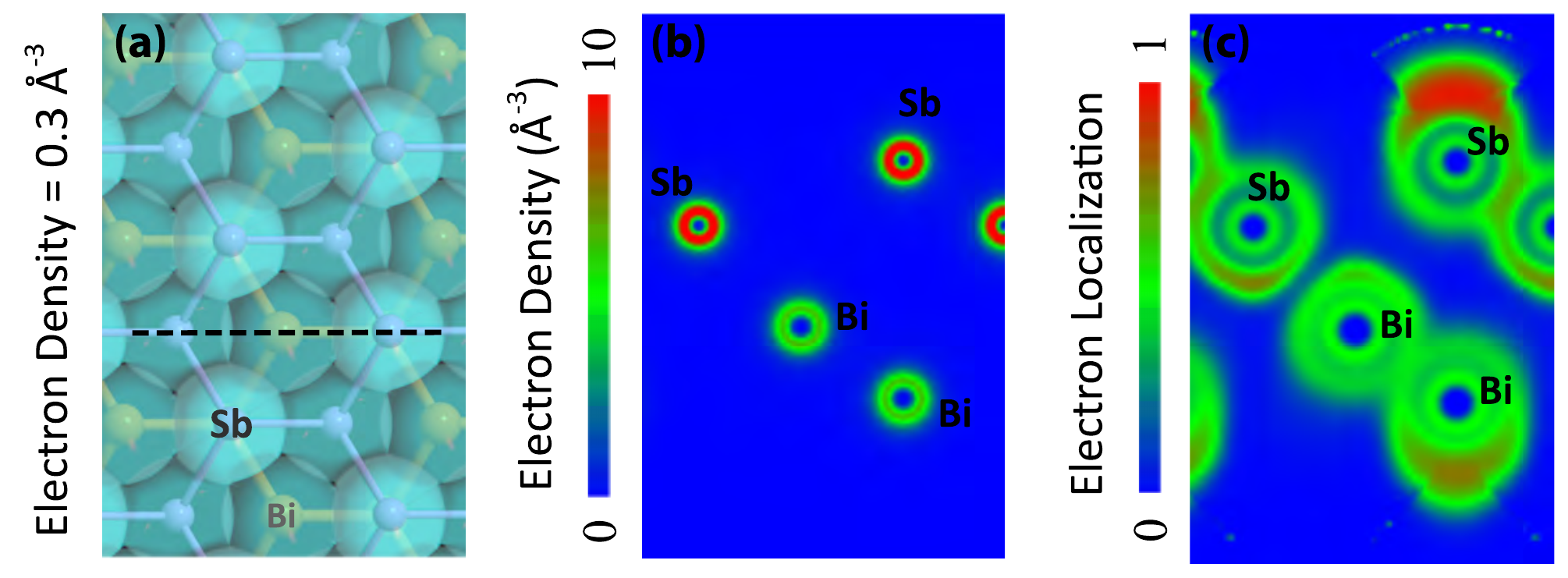}
	\caption{\textbf{(a)} Top view of the overall electron
		density
		with an iso-value
		of 0.6 \AA$^{-3} $. \textbf{(b,c)} Electron density and ELF of the
		cross
		section along
		the dashed line in (a), respectively. All three were calculated
		at
		SOGGA level of theory.}
	\label{elf}
\end{figure*}

To have further insight into the electronic properties of the
HS, we calculated the electron density and electron localization
function (ELF) \cite{becke90, savin92} (Fig.\ref{elf}). Electron density can
bring
information about how electrons are distributed in the lattice, and ELF allows
evaluating the chemical interactions from the charge localization
between individual atoms. Our calculations show that there is a little and
uniform electron density of about 0.3 {\AA}$^{-3}$ (Fig.\ref{elf}a) between
atoms and layers. Electron
density is significant around atoms and it decreases with going far from them.
The maximum electron density is collected around Sb atoms with a value of 10
{\AA}$^{-3}$
and subsequently Bi atoms have an electron density of 5 {\AA}$^{-3}$ around
them (Fig.\ref{elf}b). Considering that electron density around Sb atoms is
one time larger than around Bi atoms, it could be said that some spatial
separation of electrons and holes may occur in the HS. In other
words, most of the electrons exist in the Sb layer and most of the holes
dominate the Bi layer.
Such separation would decrease the recombinations of photo-generated electrons
and holes in solar cells and dramatically increase their efficiency.
That phenomenon has also been observed in heterostructures like InSe/Sb,
MoS\textsubscript{2}/MoSe\textsubscript{2}, and
MoS\textsubscript{2}/ReS\textsubscript{2} \cite{zhang2019, bellus17,
	ceballos14}.

It can be seen from the ELF (Fig.\ref{elf}c) that the highest localization
is above the upper Sb atoms and subsequently, under the
lower Bi atoms. There is no localization in the shortest distance between the
upper and the lower layers. Therefore, no chemical covalent
bonds exist between the layers and other mechanisms like Van-der Waals's
interactions or
orbitals hybridization may connect the layers. The lack of localization and the
absence of covalent bonds  between
the layers are also reported in the previous
heterostructures like G/Sb, PdTe\textsubscript{2}/Sb \cite{chen16, wu17} as
well as Antimonene grown on (111) Ag substrate \cite{shao18}.

\subsubsection{Appying Electric Field and Strain}
\begin{figure*}[h!]
	\centering
	\includegraphics[width=0.9\textwidth]{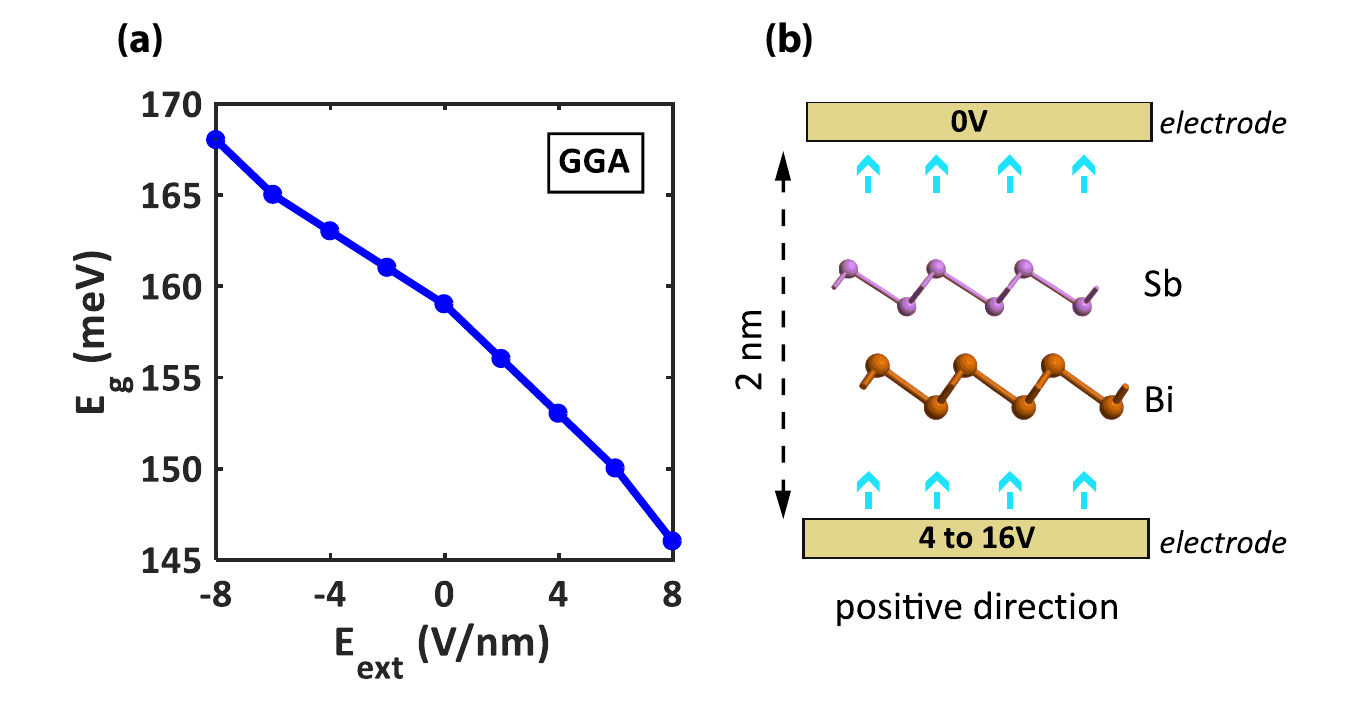}
	\caption{\textbf{(a)} Variation of the Sb/Bi HS GGA bandgap,
		by
		applying
		external electric field. \textbf{(b)} Schematic of applying external
		filed
		in the
		positive direction.}
	\label{eext}
\end{figure*}

As mentioned before, our calculations show that the Sb/Bi HS is a
semimetal at SOGGA level of theory. Existence of a bandgap in a 2D material
would bring hope for designing nano-scale field-effect
transistors (FETs). Of the methods for opening a bandgap in non-gap materials
are applying an external electric field and/or tensile and compressive strain.
For example, Li et al have tunned the bandgap of MoS\textsubscript{2}/Sb
heterostructure from 0 to 1.2 eV by applying external electric field
\cite{li19}. Lu et al have also reported that with applying tensile strain,
the bandgap of h-BN/Sb heterostructure increases \cite{lu16}. Besides, Dong at
al. could tune the band gap of As/MX\textsubscript{2} (X = S, Se \& M = Mo, W)
heterostructures from 0 to 1.5 eV by
applying an electric field from -8 to 4 V/nm (with positive direction from As
to MX\textsubscript{2}) and biaxial strain from -10 to 15$\%$
\cite{dong17}.
We also applied electric field and biaxial strain to tune the bandgap of Sb/Bi
HS. In applying the electric field, we
chose the positive direction from Bi to Sb layer and vise versa and applied the
electric field from -8 to 8 V/nm (Fig.\ref{eext}b). We also applied the
biaxial strain within
the range from -14 to +14$\%$ with steps of 2$\%$.

By use of electric field, the GGA bandgap of the HS can be tuned within the
range from 146 meV to 168 meV. It increases with the negative electric fields
and decreases with the positive ones (Fig.\ref{eext}). Also by applying tensile
strains, the bandgap encloses immediately and by applying compressive strain,
the bandgap reaches to the value of around 76 meV under the strain of
-2$\%$ and
finally encloses under the strain of -4$\%$. Our investigations at SOGGA level
of theory shows that the electronic properties of the material are robust
against the electric field and strain and can not be tuned trough these
methods.
Considering the importance of the SOC interactions in this HS, it
could be summarised that it is a semimetal in which
the  CBM and VBM touch the Fermi level. Most of the electrons are collected in
the Sb layer and most of the holes rule the Bi layer. The semimetallic
properties are robust against applying the electric field and strain and would
not change with such external factors.

\subsection{Mechanical Properties of the Sb/Bi HS}
\begin{figure*}[h!]
	\centering
z	\includegraphics[width=0.9\textwidth]{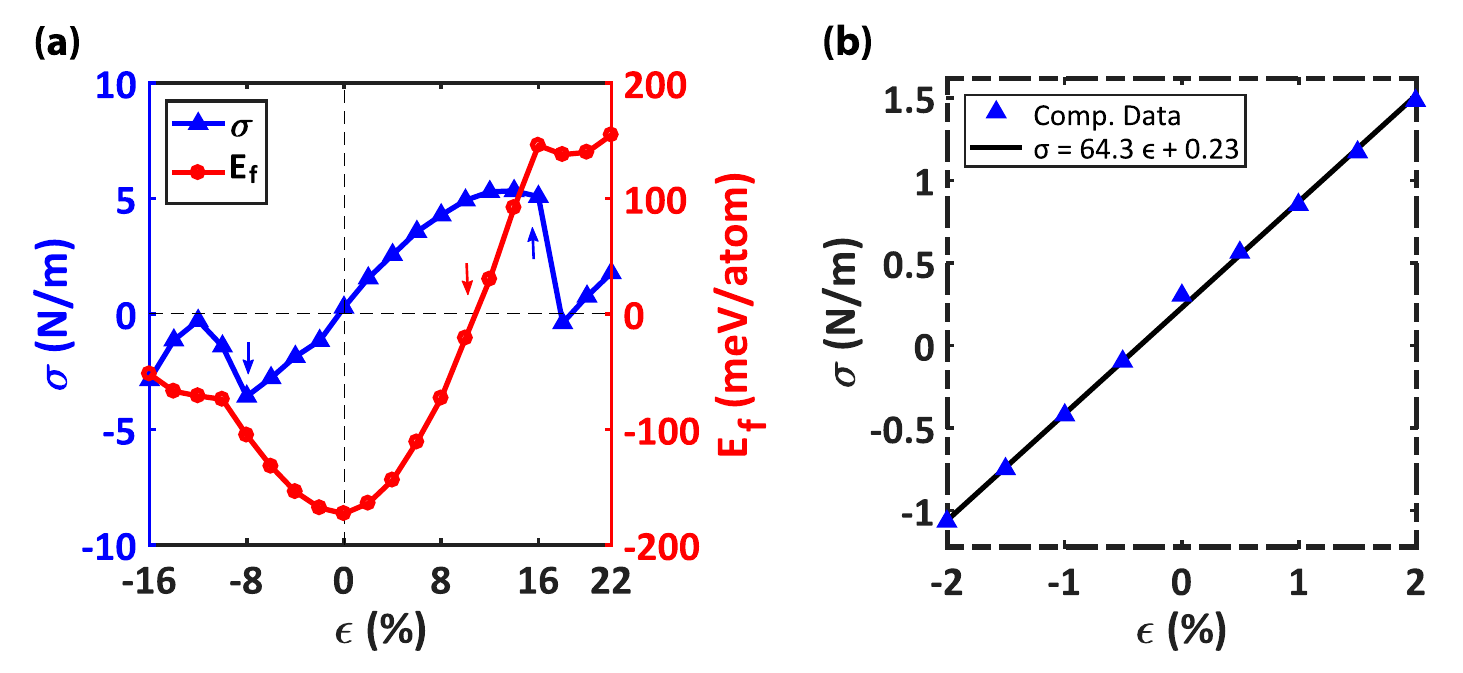}
	\caption{The Mechanical properties of Sb/Bi HS. \textbf{(a)}
		Curves
		of stress
		and
		cohesive energies variations by strain. \textbf{(b)} The stress-strain
		curve within the strain range of -2$\%$ to 2$\%$ , with a fitted
		function used for calculating Young's modulus.}
	\label{mech}
\end{figure*}

\begin{figure}[h!]
	\centering
	\includegraphics[width=0.4\textwidth]{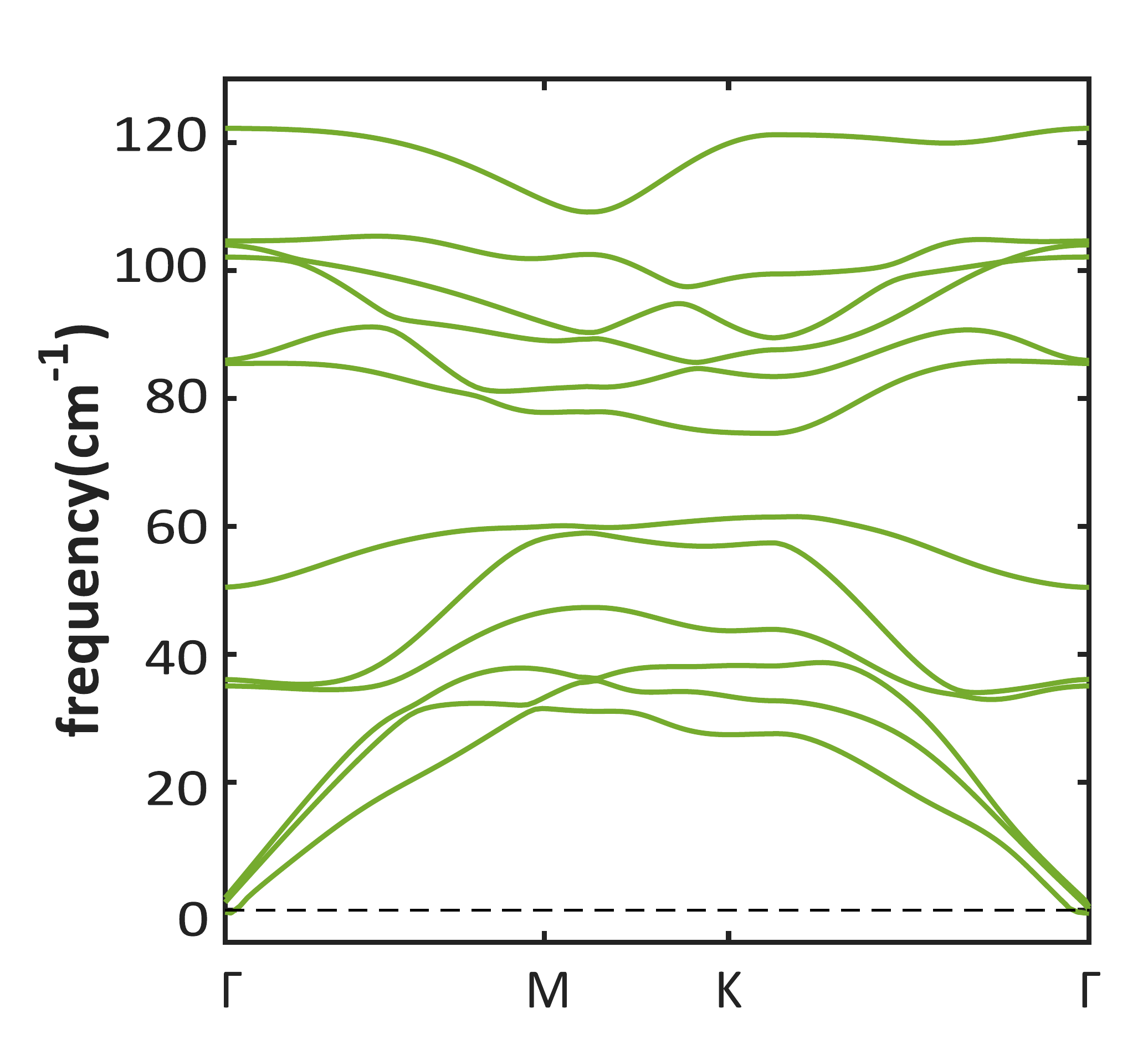}
	\caption{Phonon dispersion spectrum for the HS under the
		strain of 14$\%$. No imaginary modes are visible, therefore the
		structure
		is stable.}
	\label{phon14}
\end{figure}

With applying strain, a
material resists and stress applies to its surface. As the applied
strain increases, the stress gets greater until under a certain critical strain
the material cracks and the stress suddenly drops. The
more the so-called critical strain would be, the more the material can be
stretched or compressed.  In addition, the cohesive energy between layers tells
about how much the layers tend to be integrated and connected to each other. By
considering the stress-strain and cohesive energy-strain relationships, one can
find
out how much a material
can be stretched and/or compressed, and remain stable. Fig.\ref{mech}a shows
the stress-strain
and cohesive energy-strain curves for the Sb/Bi HS. The stress is 0.28
N/m in the pristine structure and by the increase of tensile strain, it
begins to increase slightly. Under the strain of 16$\%$, it reaches to
5.08 N/m and then drops. Therefore, the critical strain and the respective
ideal strength are 16$\%$ and 5.08 N/m, respectively.
Besides,  the cohesive energy is -173 meV/atom in the
pristine structure, which by the increase of tensile strain exponentially
increases and finally reaches to its maximum negative value of -21 meV/atom and
then gets positive. The positiveness of the cohesive energy means that
the mechanism of formation of the HS becomes endothermic and the connection
between the layers would be canceled. Moreover, the phonon dispersion
calculations show that the material is still stable under the strain of 14$\%$
(Fig.\ref{phon14}). Intuitively,
we predict that under the tensile strain of 10$\%$ the Sb and Bi layers
separate
and under the strain of 16$\%$, the material cracks.
With applying compressive strain, the stress descends towards the
negative direction and it reaches to -3.67 N/m under the strain of -8$\%$, and
after that it suddenly ascends.
Therefore, we took the -8$\%$ as the critical compressive strain with a
corresponding compressive ideal strength of -3.67 N/m.
It should be noted that the cohesive energy is negative within this range,
that is, the layers are connected to each other.

For further investigations about the mechanical properties, we calculated
Young's
modulus, which is defined by the equations below,
\begin{equation}\label{youngeq}
E=\frac{\sigma}{\epsilon}
\end{equation}
where, $\epsilon$ and $\sigma$ are the strain and stress, respectively.
According
to the equation (\ref{youngeq}), the more Young's modulus is, the greater the
response of the
material to the applied strain would be. That is, the material resists more
against and has less intention for being stretched or compressed.
To calculate Young's modulus, we applied strains within the range of -2$\%$ to
2$\%$ with steps of 0.5$\%$. The variations of the stress in this range of
strain is approximately linear. Furthermore, we fitted the computational data
with a linear function (Fig.\ref{mech}b). By calculating the gradiant of the
linear fitted function, we anticipate Young's modulus of 64.3$\%$ N/m for the
HS.

We compared our mechanical results about the HS with other 2D materials
including
Arsenene, Antimonene, Silicene, Graphene, and Molibdanium disulfide
(MoS\textsubscript{2}) (Table \ref{mechtab}). In comparison with group V 2D
materials, the
critical strain of the HS is bigger than Phosphorene's but smaller than
Antimonene's and Arsenene's. Interestingly, Young's modulus of the HS is
greater than all of the group V 2D materials' and even than Silicene's.
Therefore, although the HS loses its stability in strains longer than 10$\%$,
its high Young's modulus do not allow for size change so easily.
\begin{table}[h]
	\caption{The mechanical properties of the Sb/Bi HS
		including
		critical strain ($ \epsilon^* $), ideal strength ($\sigma^*$), and
		Young's modulus ($E_{xy}$) compared with other 2D materials.}
	\label{mechtab}
	
	\begin{tabular}{llll}
		\hline\hline
		2D material & $\epsilon^* (\%)$ & $\sigma^* (N/m)$ & $E_{xy}
		(N/m)$\\
		\hline
		Phosphorene & 8 \cite{wei14} & 20.26 \cite{wei14} & 23
		\cite{zhao14}\\
		Antimonene \cite{liu19} & 20 & 3.75 & 32.9 \\
		Arsenene \cite{liu19} & 17 & 5.15 & 52.8 \\
		Bismuthene\cite{aghdasi19} &15 & n/a &26.25\\
		Silicene \cite{mortazavi17} & 17.5 & 7.20 & 61.7 \\
		{Sb/Bi HS (this study}) & {16} & {5.08} &
		{64.3} \\
		MoS\textsubscript{2}  & 28\cite{li12} & 15.73\cite{li12}
		& 118 \cite{lorenz12} \\
		Graphene  & 19 \cite{liu07} & 32.93 \cite{liu07} & 345
		\cite{kudin01} \\
		\hline\hline
	\end{tabular}
\end{table}
\subsection{Optical Properties of the Sb/Bi HS}
The optical properties are described by the complex dielectric
function, $\epsilon(\omega)=\epsilon_1(\omega)+i\epsilon_2(\omega)$.
We first calculated the susceptibility tensor by Kubo-Greenwood formula
\cite{martin04},

\begin{equation}
\chi_{ij}(\omega)=-\frac{e^2\hbar^4}{m^2\epsilon_0V\omega^2}
\sum\frac{f(E_m)-f(E_n)}{E_{nm}-\hbar\omega-i\Gamma}\pi_{nm}^i\pi_{mn}^j
\end{equation}

\noindent where $\pi_{nm}^i$ is the i-th component of the dipole matrix
between state n and m, V the volume, $\Gamma$ the broadening, and f the Fermi
function. Subsequently, one may calculate the dielectric function by the
equation below,

\begin{equation}
\epsilon(\omega)=1+\chi(\omega)
\end{equation}

\noindent Furthermore, the optical absorption coefficient ($\alpha$) can be
calculated
through,

\begin{equation}
\alpha=2\frac{\omega}{c}\sqrt{\frac{\sqrt{\epsilon_1^2+\epsilon_2^2}
		-\epsilon_1}{2}}
\end{equation}

\noindent where c is the speed of light.

\begin{figure}[h!]
	\centering
	\includegraphics[width=0.4\textwidth]{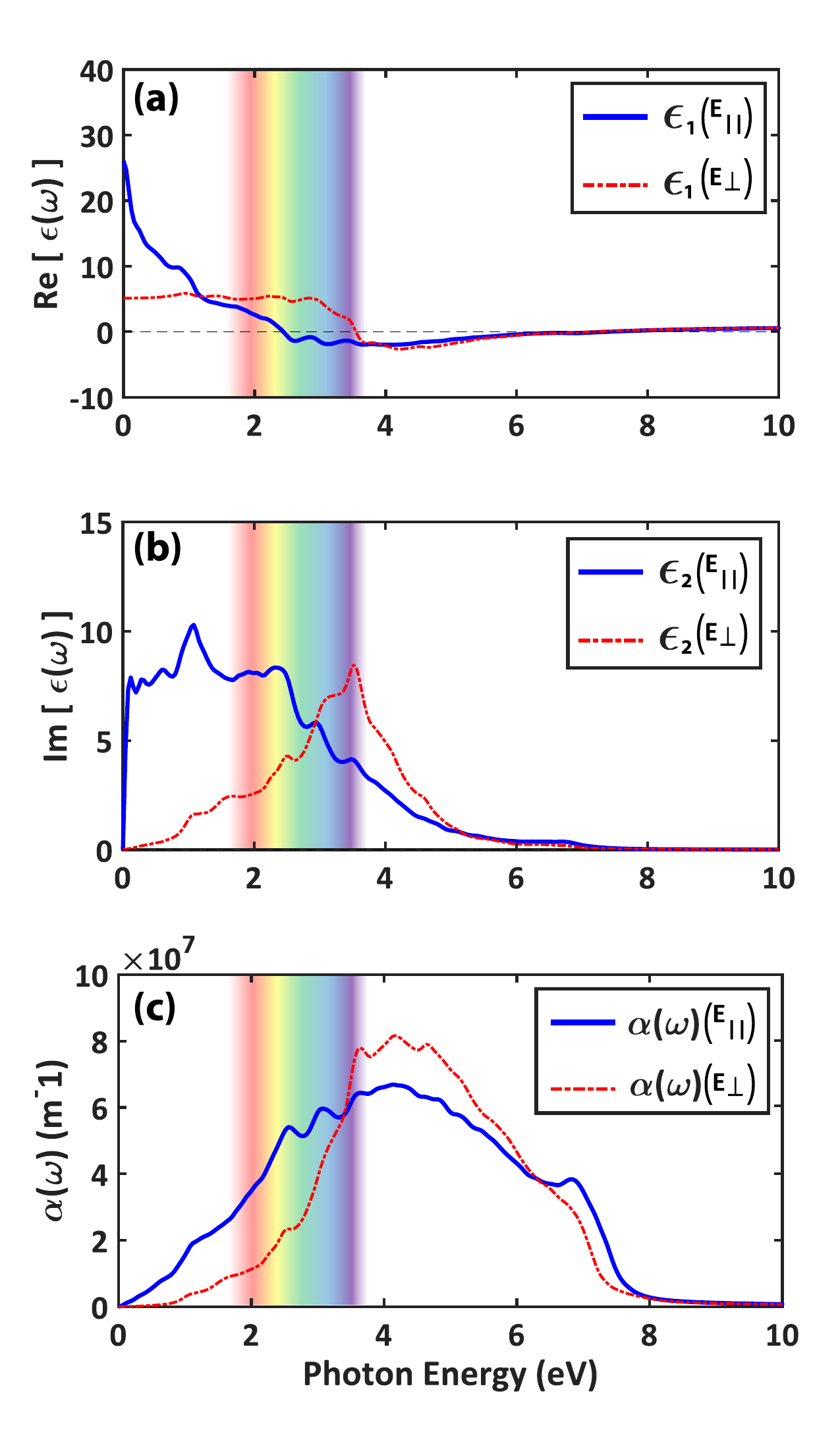}
	\caption{Optical properties of Sb/Bi HS. \textbf{(a)} Real and
		\textbf{(b)}
		imaginary parts of the dielectric function. \textbf{(c)} Optical
		absorption
		coefficient.}
	\label{opt}
\end{figure}

Fig.\ref{opt} shows the real and imaginary parts of the dielectric function
as well as the optical absorption of the Sb/Bi HS. In these calculations, we
used the GGA  with considerations of the SOC interactions (SOGGA). For the
isotropy of the
crystalline
structure of the HS in the xy-plane, the dielectric functions for linear
polarization along x and y direction are similar ($\epsilon^{xx} =
\epsilon^{yy}$). Therefore we note the optical properties for the parallel
($E_{\parallel}$, along with the x- and y-direction) and perpendicular
($E_{\perp}$, along with the z-direction) polarization regarding the
sheet plane.

As can be seen in Fig.\ref{opt}, the behavior of the HS is different for
different polarization. To the best of our knowledge, the negative values of
the real part of
the dielectric function show a metallic optical properties. As can be seen from
Fig.
\ref{opt}a there are negative values for the real part of the dielectric
function in the UV spectrum which stands for metallic properties. The
beginning of the metallic properties differs in different polarization. The HS
begins to be metallic from 3.56 eV (348 nm) in the UV and 2.44 eV (548 nm) in
the visible region for the perpendicular and parallel polarization,
respectively. The metallic characteristics are also reported for monolayer
Antimonene \cite{singh16, xu17} but unlike the Sb/Bi HS, it begins in the UV
region, not visible.

\begin{figure}
	\centering
	\includegraphics[width=0.4\textwidth]{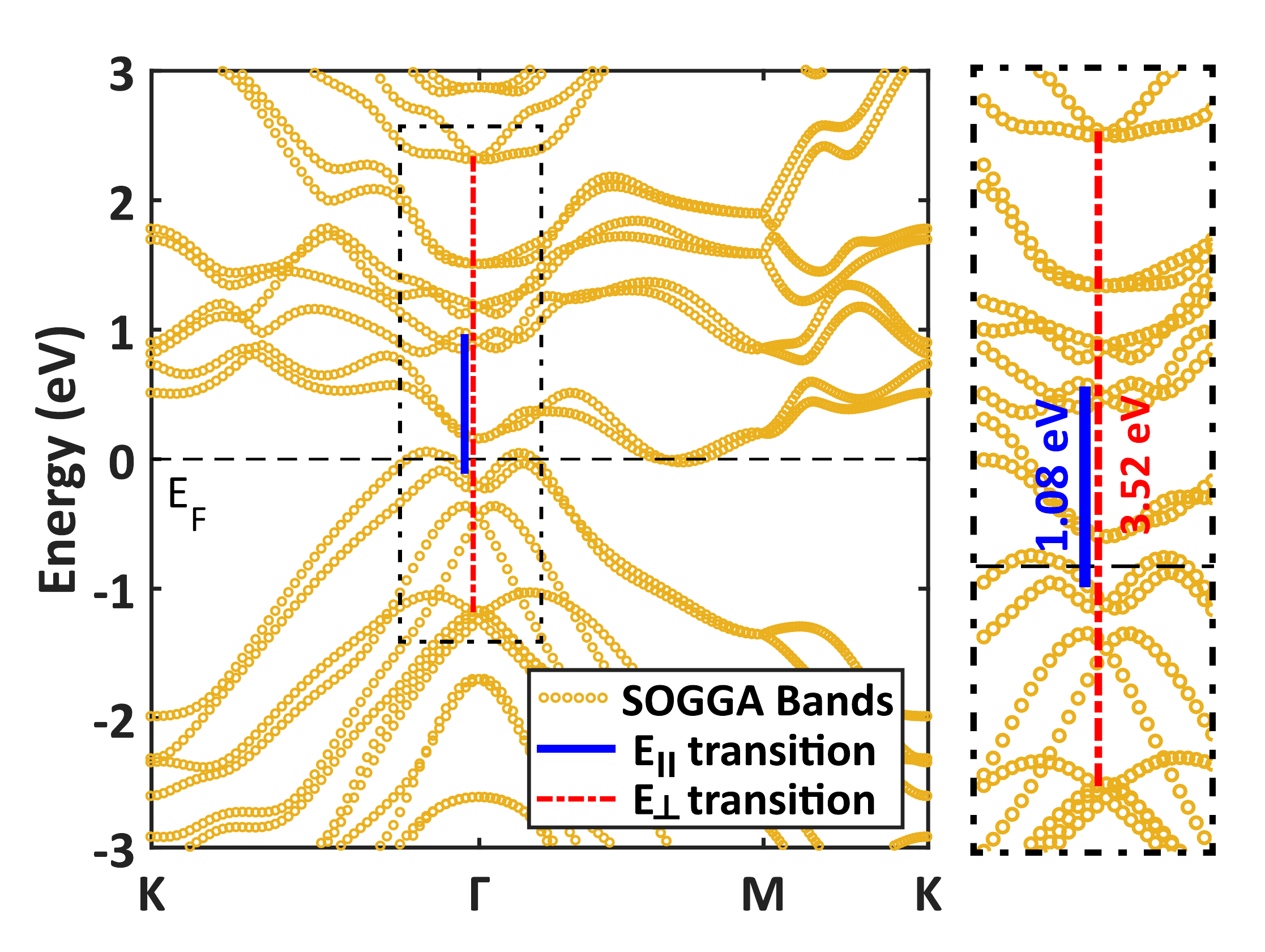}
	\caption{Direct electron transitions between bands below and
		above the Fermi level, for parallel (1.08 eV) and perpendicular (3.52
		eV) polarization.}
	\label{trans}
\end{figure}

Peaks in the imaginary part of the dielectric function are due to
absorption of the incident photons and direct transitions of the electrons
between bands below and above the Fermi energy \cite{mohan12}. As can be seen
from Fig.\ref{opt}b there are two major peaks in the imaginary part of
the
dielectric function in 1.08 and 3.52 eV, respectively. By analyzing
the band structure and precise measuring of the distance between bands below
and above the Fermi level in different k points, one can make a relative
assessment about how the electron transitions take place. Our analysis shows
that these transitions for perpendicular and parallel polarization take place
at the $\Gamma$ point, where is the only point which the
energy difference between some conduction and valance bands are exactly equal
to transition energies (Fig.\ref{trans}).
Direct electron transitions are also
reported for the monolayer Antimonene in 3.4, 4.7, 5.9, 6.9, and 8.0 eV at
different k points
\cite{singh16}.

Fig.\ref{opt}c shows the absorption coefficient for the Sb/Bi HS. In the
infrared region, the absorption is not so much, but it is a bit more than the
monolayer Antimonene \cite{singh16}. Therefore, compared with the
monolayer Antimonene, the HS
absorbs more heat. The parallel and perpendicular polarization are absorbed
deeper in the visible and UV regions, respectively. The maximum of the
absorption is about $8 \times10^7$ for photon energy of $\sim$4 eV (310 nm) in
the UV region. Our
investigations show
that the absorption in the visible region for the HS is sort of 4 times
greater than the monolayer Antimonene, in which it is reported neglectable
\cite{xu17}. Such increase in the absorption in the visible region is also
reported in Sb/GaAs heterostructure \cite{wang17}.

In general, the dependence of reflectance and absorption on the polarization of
the incident light, along with the increase of absorption of perpendicular
polarization in the visible region can improve the applications of the Sb/Bi HS
in beam splitters and nano-scale mirrors compared with the monolayer Antimonene.

\section{Conclusion}

In summary, we systematically calculated the structural, electronic,
mechanical, and optical properties of the Sb/Bi heterostructure using the first
principle density functional theory. Our investigations show that this
heterostructure is theoretically stable and possible,
and the ABi model is the most stable stacking. Its
semimetallic electronic characteristics are robust against external
factors such as electric field and mechanical strain. The Sb/Bi
heterostructure is mechanically a stiff material with Young's modulus of 64.3
N/m which is larger than all known group V monolayers' and even
than Silicene's. The optical properties of the HS are dependant on the
polarization
of the incident light.
It begins to be metallic in the visible and UV region for parallel and
perpendicular polarization, respectively. In comparison with monolayer
Antimonene, the absorption of the perpendicular polarization in the visible
region is improved. The heterostructure has a good optical absorption and
reflectance in the visible and UV region. In addition, the polarization of the
incident light makes significant difference in the optical propertie. Therefore
we predict its potential applications in beam splitters and nano-scale mirrors.

\section*{Acknowledgment}
We acknowledge Mohammad Ali Mohebpour and Salimeh Mahdavifar for helpful
discussions and recommendations. We are thankful to the Research Council of the
University of Guilan for the partial support of this research.

\section*{Declaration of Interests}
The authors declare that they have no known competing financial interests or
personal relationships that could have appeared to influence the work reported
in this paper.

\bibliography{abbr}

\begin{thebibliography}{10}
\expandafter\ifx\csname url\endcsname\relax
  \def\url#1{\texttt{#1}}\fi
\expandafter\ifx\csname urlprefix\endcsname\relax\def\urlprefix{URL }\fi
\expandafter\ifx\csname href\endcsname\relax
  \def\href#1#2{#2} \def\path#1{#1}\fi

\bibitem{lu14}
W.~Lu, H.~Nan, J.~Hong, Y.~Chen, C.~Zhu, Z.~Liang, X.~Ma, Z.~Ni, C.~Jin,
  Z.~Zhang, Plasma-assisted fabrication of monolayer phosphorene and its raman
  characterization, Nano Res. 7~(6) (2014) 853--859.

\bibitem{frank07}
I.~W. Frank, D.~M. Tanenbaum, A.~M. van~der Zande, P.~L. McEuen, Mechanical
  properties of suspended graphene sheets, J. Vac. Sci. Technol. B 25~(6)
  (2007) 2558--2561.

\bibitem{novoselov2012roadmap}
K.~S. Novoselov, V.~Fal, L.~Colombo, P.~Gellert, M.~Schwab, K.~Kim, et~al., A
  roadmap for graphene, nature 490~(7419) (2012) 192.

\bibitem{novoselov04}
K.~S. Novoselov, A.~K. Geim, S.~V. Morozov, D.~Jiang, Y.~Zhang, S.~V. Dubonos,
  I.~V. Grigorieva, A.~A. Firsov, Electric field effect in atomically thin
  carbon films, Science 306~(5696) (2004) 666--669.

\bibitem{geim11}
A.~K. Geim, Random walk to graphene (nobel lecture), Angew. Chem. Int. Ed.
  50~(31) (2011) 6966--6985.

\bibitem{kamal15}
C.~Kamal, M.~Ezawa, Arsenene: Two-dimensional buckled and puckered honeycomb
  arsenic systems, Phys. Rev. B 91~(8) (2015) 085423.

\bibitem{zhang2015}
S.~Zhang, Z.~Yan, Y.~Li, Z.~Chen, H.~Zeng, Atomically thin arsenene and
  antimonene: semimetal–semiconductor and indirect–direct band‐gap
  transitions, Angew. Chem. Int. Ed. 54~(10) (2015) 3112--3115.

\bibitem{ni11}
Z.~Ni, Q.~Liu, K.~Tang, J.~Zheng, J.~Zhou, R.~Qin, Z.~Gao, D.~Yu, J.~Lu,
  Tunable bandgap in silicene and germanene, Nano Lett. 12~(1) (2011) 113--118.

\bibitem{davila14}
M.~E. Dávila, L.~Xian, S.~Cahangirov, A.~Rubio, G.~Le~Lay, Germanene: a novel
  two-dimensional germanium allotrope akin to graphene and silicene, New J.
  Phys. 16~(9) (2014) 095002.

\bibitem{chhowalla13}
M.~Chhowalla, H.~S. Shin, G.~Eda, L.-J. Li, K.~P. Loh, H.~Zhang, The chemistry
  of two-dimensional layered transition metal dichalcogenide nanosheets, Nat.
  Chem. 5~(4) (2013) 263.

\bibitem{liu03}
L.~Liu, Y.~P. Feng, Z.~X. Shen, Structural and electronic properties of h-bn,
  Phys. Rev. B 68~(10) (2003) 104102.

\bibitem{bafekry2019}
A.~Bafekry, M.~Ghergherehchi, S.~F. Shayesteh, Tuning the electronic and
  magnetic properties of antimonene nanosheets via point defects and external
  fields: first-principles calculations, Phys. Chem. Chem. Phys. 21~(20) (2019)
  10552--10566.

\bibitem{mohebpour2018}
M.~A. Mohebpour, M.~Saffari, H.~R. Soleimani, M.~B. Tagani, High performance of
  mixed halide perovskite solar cells: Role of halogen atom and plasmonic
  nanoparticles on the ideal current density of cell, Physica E 97 (2018)
  282--289.

\bibitem{xu2018two}
Y.~Xu, C.-Y. Hsieh, L.~Wu, L.~Ang, Two-dimensional transition metal
  dichalcogenides mediated long range surface plasmon resonance biosensors, J.
  Phys. D 52~(6) (2018) 065101.

\bibitem{reddy2011graphene}
D.~Reddy, L.~F. Register, G.~D. Carpenter, S.~K. Banerjee, Graphene
  field-effect transistors, J. Phys. D 44~(31) (2011) 313001.

\bibitem{erchak01}
A.~A. Erchak, D.~J. Ripin, S.~Fan, P.~Rakich, J.~D. Joannopoulos, E.~P. Ippen,
  G.~S. Petrich, L.~A. Kolodziejski, Enhanced coupling to vertical radiation
  using a two-dimensional photonic crystal in a semiconductor light-emitting
  diode, Appl. Phys. Lett. 78~(5) (2001) 563--565.

\bibitem{saffari2017dft}
M.~Saffari, M.~A. Mohebpour, H.~R. Soleimani, M.~B. Tagani, Dft analysis and
  fdtd simulation of ch3nh3pbi3- x cl x mixed halide perovskite solar cells:
  role of halide mixing and light trapping technique, J. Phys. D 50~(41) (2017)
  415501.

\bibitem{cai10}
Y.~Cai, S.~Liu, X.~Yin, Q.~Hao, M.~Zhang, T.~Wang, Facile preparation of porous
  one-dimensional mn2o3 nanostructures and their application as anode materials
  for lithium-ion batteries, Physica E 43~(1) (2010) 70--75.

\bibitem{wang15}
G.~Wang, R.~Pandey, S.~P. Karna, Atomically thin group v elemental films:
  theoretical investigations of antimonene allotropes, ACS Appl. Mater.
  Interfaces 7~(21) (2015) 11490--6.
\newblock \href {http://dx.doi.org/10.1021/acsami.5b02441}
  {\path{doi:10.1021/acsami.5b02441}}.

\bibitem{akturk15}
O.~z. Aktürk, V.~O. Özçelik, S.~Ciraci, Single-layer crystalline phases of
  antimony: Antimonenes, Phys. Rev. B 91~(23) (2015) 235446.

\bibitem{akturk16}
E.~Aktürk, O.~z. Aktürk, S.~Ciraci, Single and bilayer bismuthene: Stability
  at high temperature and mechanical and electronic properties, Phys. Rev. B
  94~(1) (2016) 014115.

\bibitem{gibaja16}
C.~Gibaja, D.~Rodriguez‐San‐Miguel, P.~Ares, J.~Gómez‐Herrero,
  M.~Varela, R.~Gillen, J.~Maultzsch, F.~Hauke, A.~Hirsch, G.~Abellán,
  Few‐layer antimonene by liquid‐phase exfoliation, Angew. Chem. Int. Ed.
  55~(46) (2016) 14345--14349.

\bibitem{singh16}
D.~Singh, S.~K. Gupta, Y.~Sonvane, I.~Lukačević, Antimonene: a monolayer
  material for ultraviolet optical nanodevices, J. Mater. Chem. C 4~(26) (2016)
  6386--6390.

\bibitem{xu17}
Y.~Xu, B.~Peng, H.~Zhang, H.~Shao, R.~Zhang, H.~Zhu, First‐principle
  calculations of optical properties of monolayer arsenene and antimonene
  allotropes, Ann. Phys. 529~(4) (2017) 1600152.

\bibitem{guo2018biaxial}
S.-D. Guo, J.~Dong, Biaxial tensile strain tuned up-and-down behavior on
  lattice thermal conductivity in $\beta$-asp monolayer, J. Phys. D 51~(26)
  (2018) 265307.

\bibitem{lu16}
H.~Lu, J.~Gao, Z.~Hu, X.~Shao, Biaxial strain effect on electronic structure
  tuning in antimonene-based van der waals heterostructures, RSC Adv. 6~(104)
  (2016) 102724--102732.
\newblock \href {http://dx.doi.org/10.1039/c6ra21781h}
  {\path{doi:10.1039/c6ra21781h}}.

\bibitem{tsai14}
M.-L. Tsai, S.-H. Su, J.-K. Chang, D.-S. Tsai, C.-H. Chen, C.-I. Wu, L.-J. Li,
  L.-J. Chen, J.-H. He, Monolayer mos2 heterojunction solar cells, ACS Nano
  8~(8) (2014) 8317--8322.

\bibitem{chen16}
X.~Chen, Q.~Yang, R.~Meng, J.~Jiang, Q.~Liang, C.~Tan, X.~Sun, The electronic
  and optical properties of novel germanene and antimonene heterostructures, J.
  Mater. Chem. C 4~(23) (2016) 5434--5441.
\newblock \href {http://dx.doi.org/10.1039/c6tc01141a}
  {\path{doi:10.1039/c6tc01141a}}.

\bibitem{soler2002siesta}
J.~M. Soler, E.~Artacho, J.~D. Gale, A.~Garc{\'\i}a, J.~Junquera,
  P.~Ordej{\'o}n, D.~S{\'a}nchez-Portal, The siesta method for ab initio
  order-n materials simulation, J. Phys. Condens. Matter 14~(11) (2002) 2745.

\bibitem{perdew97}
J.~P. Perdew, K.~Burke, M.~Ernzerhof, Generalized gradient approximation made
  simple, Phys. Rev. Lett. 78~(7) (1997) 1396.
\newblock \href {http://dx.doi.org/10.1103/PhysRevLett.78.1396}
  {\path{doi:10.1103/PhysRevLett.78.1396}}.

\bibitem{grimme04}
S.~Grimme, Accurate description of van der waals complexes by density
  functional theory including empirical corrections, J. Comput. Chem. 25~(12)
  (2004) 1463--1473.

\bibitem{mounet18}
N.~Mounet, M.~Gibertini, P.~Schwaller, D.~Campi, A.~Merkys, A.~Marrazzo,
  T.~Sohier, I.~E. Castelli, A.~Cepellotti, G.~Pizzi, N.~Marzari,
  Two-dimensional materials from high-throughput computational exfoliation of
  experimentally known compounds, Nat. Nanotechnol. 13~(3) (2018) 246--252.
\newblock \href {http://dx.doi.org/10.1038/s41565-017-0035-5}
  {\path{doi:10.1038/s41565-017-0035-5}}.

\bibitem{zhang16}
S.~Zhang, M.~Xie, F.~Li, Z.~Yan, Y.~Li, E.~Kan, W.~Liu, Z.~Chen, H.~Zeng,
  Semiconducting group 15 monolayers: a broad range of band gaps and high
  carrier mobilities, Angew. Chem. Int. Ed. 55~(5) (2016) 1666--1669.

\bibitem{liu17}
M.-Y. Liu, Y.~Huang, Q.-Y. Chen, Z.-Y. Li, C.~Cao, Y.~He, Strain and electric
  field tunable electronic structure of buckled bismuthene, RSC Adv. 7~(63)
  (2017) 39546--39555.

\bibitem{liu19}
G.~Liu, Z.~Gao, J.~Zhou, Strain effects on the mechanical properties of group-v
  monolayers with buckled honeycomb structures, Physica E 112 (2019) 59--65.

\bibitem{zhao15}
M.~Zhao, X.~Zhang, L.~Li, Strain-driven band inversion and topological aspects
  in antimonene, Sci. Rep. 5 (2015) 16108.

\bibitem{manzeli20172d}
S.~Manzeli, D.~Ovchinnikov, D.~Pasquier, O.~V. Yazyev, A.~Kis, 2d transition
  metal dichalcogenides, Nat. Rev. Mater.s 2~(8) (2017) 17033.

\bibitem{ezawa2017triplet}
M.~Ezawa, Triplet fermions and dirac fermions in borophene, Phys. Rev. B 96~(3)
  (2017) 035425.

\bibitem{becke90}
A.~D. Becke, K.~E. Edgecombe, A simple measure of electron localization in
  atomic and molecular systems, J. Chem. Phys. 92~(9) (1990) 5397--5403.

\bibitem{savin92}
A.~Savin, O.~Jepsen, J.~Flad, O.~K. Andersen, H.~Preuss, H.~G. von Schnering,
  Electron localization in solid‐state structures of the elements: the
  diamond structure, Angew. Chem. Int. Ed. 31~(2) (1992) 187--188.

\bibitem{zhang2019}
Z.~Zhang, Y.~Zhang, Z.~Xie, X.~Wei, T.~Guo, J.~Fan, L.~Ni, Y.~Tian, J.~Liu,
  L.~Duan, Tunable electronic properties of an sb/inse van der waals
  heterostructure by electric field effects, PCCP 21~(10) (2019) 5627--5633.

\bibitem{bellus17}
M.~Z. Bellus, M.~Li, S.~D. Lane, F.~Ceballos, Q.~Cui, X.~C. Zeng, H.~Zhao,
  Type-i van der waals heterostructure formed by mos 2 and res 2 monolayers,
  Nanoscale Horiz. 2~(1) (2017) 31--36.

\bibitem{ceballos14}
F.~Ceballos, M.~Z. Bellus, H.-Y. Chiu, H.~Zhao, Ultrafast charge separation and
  indirect exciton formation in a mos2–mose2 van der waals heterostructure,
  ACS Nano 8~(12) (2014) 12717--12724.

\bibitem{wu17}
X.~Wu, Y.~Shao, H.~Liu, Z.~Feng, Y.~Wang, J.~Sun, C.~Liu, J.~Wang, Z.~Liu,
  S.~Zhu, Epitaxial growth and air‐stability of monolayer antimonene on
  pdte2, Adv. Mater. 29~(11) (2017) 1605407.

\bibitem{shao18}
Y.~Shao, Z.-L. Liu, C.~Cheng, X.~Wu, H.~Liu, C.~Liu, J.-O. Wang, S.-Y. Zhu,
  Y.-Q. Wang, D.-X. Shi, Epitaxial growth of flat antimonene monolayer: A new
  honeycomb analogue of graphene, Nano Lett. 18~(3) (2018) 2133--2139.

\bibitem{li19}
W.~Li, Y.~Ma, X.~Wang, X.~Dai, Electric field effects on the electronic
  structures of mos2/antimonene van der waals heterostructure, Solid State
  Commun. 293 (2019) 28--32.

\bibitem{dong17}
M.~M. Dong, C.~He, W.~X. Zhang, Tunable electronic properties of arsenene and
  transition-metal dichalcogenide heterostructures: a first-principles
  calculation, J. Phys. Chem. C 121~(40) (2017) 22040--22048.

\bibitem{wei14}
Q.~Wei, X.~Peng, Superior mechanical flexibility of phosphorene and few-layer
  black phosphorus, Appl. Phys. Lett. 104~(25) (2014) 251915.

\bibitem{zhao14}
Y.~Zhao, A.~M. Nardes, K.~Zhu, Solid-state mesostructured perovskite ch3nh3pbi3
  solar cells: charge transport, recombination, and diffusion length, J. Phys.
  Chem. Lett. 5~(3) (2014) 490--494.

\bibitem{aghdasi19}
P.~Aghdasi, R.~Ansari, S.~Rouhi, M.~Goli, On the elastic and plastic properties
  of the bismuthene adsorbed by h, f, cl and br atoms, Superlattice. Microst.
  (2019) 106242.

\bibitem{mortazavi17}
B.~Mortazavi, O.~Rahaman, M.~Makaremi, A.~Dianat, G.~Cuniberti, T.~Rabczuk,
  First-principles investigation of mechanical properties of silicene,
  germanene and stanene, Physica E 87 (2017) 228--232.

\bibitem{li12}
T.~Li, Ideal strength and phonon instability in single-layer mos 2, Phys. Rev.
  B 85~(23) (2012) 235407.

\bibitem{lorenz12}
T.~Lorenz, D.~Teich, J.-O. Joswig, G.~Seifert, Theoretical study of the
  mechanical behavior of individual tis2 and mos2 nanotubes, J. Phys. Chem. C
  116~(21) (2012) 11714--11721.

\bibitem{liu07}
F.~Liu, P.~Ming, J.~Li, Ab initio calculation of ideal strength and phonon
  instability of graphene under tension, Phys. Rev. B 76~(6) (2007) 064120.

\bibitem{kudin01}
K.~N. Kudin, G.~E. Scuseria, B.~I. Yakobson, C 2 f, bn, and c nanoshell
  elasticity from ab initio computations, Phys. Rev. B 64~(23) (2001) 235406.

\bibitem{martin04}
R.~M. Martin, Electronic structure: basic theory and practical methods,
  Cambridge university press, 2004.

\bibitem{mohan12}
B.~Mohan, A.~Kumar, P.~K. Ahluwalia, A first principle study of interband
  transitions and electron energy loss in mono and bilayer graphene: Effect of
  external electric field, Physica E 44~(7-8) (2012) 1670--1674.

\bibitem{wang17}
N.~Wang, D.~Cao, J.~Wang, P.~Liang, X.~Chen, H.~Shu, Interface effect on
  electronic and optical properties of antimonene/gaas van der waals
  heterostructures, J. Mater. Chem. C 5~(37) (2017) 9687--9693.

\end{thebibliography}

\end{document}